\pdfoutput=1

\documentclass[11pt,twoside,a4paper,cmspaper,final,collab]{cms-tdr}
\def\svnVersion{382821}\def\cmsCernNoTag{CERN-EP/2016-169}\def\cmsCernDate{\today}\def\cmsMessage{Published in Physics Letters B as \href{http://dx.doi.org/10.1016/j.physletb.2017.01.011}{\doi{10.1016/j.physletb.2017.01.011.}}}

\begin{document}\cmsNoteHeader{SMP-16-002}

\hyphenation{had-ron-i-za-tion}
\hyphenation{cal-or-i-me-ter}
\hyphenation{de-vices}

\RCS$Revision: 382785 $
\RCS$HeadURL: svn+ssh://svn.cern.ch/reps/tdr2/papers/SMP-16-002/trunk/SMP-16-002.tex $
\RCS$Id: SMP-16-002.tex 382785 2017-01-25 16:04:35Z dntaylor $
\newlength\cmsFigWidth
\ifthenelse{\boolean{cms@external}}{\setlength\cmsFigWidth{0.85\columnwidth}}{\setlength\cmsFigWidth{0.4\textwidth}}
\ifthenelse{\boolean{cms@external}}{\providecommand{\cmsLeft}{top\xspace}}{\providecommand{\cmsLeft}{left\xspace}}
\ifthenelse{\boolean{cms@external}}{\providecommand{\cmsRight}{bottom\xspace}}{\providecommand{\cmsRight}{right\xspace}}
\newcommand{\ZZ}{\ensuremath{\cPZ\cPZ}\xspace}
\newcommand{\WZ}{\ensuremath{\PW\cPZ}\xspace}
\newcommand{\pp}{\ensuremath{\Pp\Pp}\xspace}
\newcommand{\mme}{\ensuremath{\Pgm\Pgm\Pe}\xspace}
\newcommand{\eee}{\ensuremath{\Pe\Pe\Pe}\xspace}
\newcommand{\eem}{\ensuremath{\Pe\Pe\Pgm}\xspace}
\newcommand{\mmm}{\ensuremath{\Pgm\Pgm\Pgm}\xspace}
\newcommand{\MATRIX}{\textsc{matrix}\xspace}
\title{Measurement of the WZ production cross section in pp collisions at $\sqrt{s} = 13$\TeV}

\author{CMS Collaboration}

\date{\today}

\abstract{
The WZ production cross section in proton-proton collisions
at $\sqrt{s} = 13$\TeV
is measured
with the CMS experiment at the LHC
using a data sample corresponding to an integrated luminosity of 2.3\fbinv.
The measurement is performed in the leptonic decay modes $\WZ\to \ell\nu\ell'\ell'$,
where $\ell, \ell' = \Pe$, $\PGm$.
The measured cross section for the range $60 < m_{{\ell'\ell'}} < 120\GeV$ is
$\sigma(\pp\to \WZ) = 39.9 \pm 3.2\stat~{}^{+2.9}_{-3.1 }\syst\pm 0.4\thy\pm 1.3\lum\unit{pb}$,
consistent with the standard model prediction.
}

\hypersetup{%
pdfauthor={CMS Collaboration},%
pdftitle={Measurement of the WZ production cross section in pp collisions at sqrt(s) = 13 TeV},%
pdfsubject={CMS},%
pdfkeywords={CMS, physics, electroweak}}

\maketitle

\section{Introduction}
\label{sec:introduction}

Measurements of the cross sections for massive gauge boson pair production in proton-proton collisions provide an essential test of the
electroweak sector of the standard model (SM).  The electroweak interaction in the SM is
determined by the non-Abelian $SU(2)_\mathrm{L} \times U(1)_\mathrm{Y}$ gauge group.  The non-Abelian nature of the electroweak gauge group
leads to gauge boson self-interactions via triple gauge couplings (TGCs) and quartic gauge couplings (QGCs).
The weak gauge boson pair production includes TGC interactions
as well as QGC interactions via vector boson scattering.
Thus, the study of diboson production can directly
test both the weak interaction and the non-Abelian nature of the electroweak gauge group.
The next-to-leading order (NLO) and next-to-next-to-leading order (NNLO) perturbative QCD corrections
for the boson pair production have substantial impact on the predicted cross sections
due to the addition of the gluon-initiated processes that are enhanced at energies available at the CERN LHC.
The increase in the cross section is significant compared to the experimental uncertainties,
allowing LHC boson pair cross section measurements to directly validate higher-order perturbative QCD calculations.

The observation of WZ  production in proton-antiproton
collisions at the Tevatron collider was reported by the CDF~\cite{Abulencia:2007tu,Aaltonen:2012vu} and
D0~\cite{Abazov:2012cj}  experiments.
The WZ production cross section in proton-proton collisions has been measured at the LHC
by the CMS experiment at $\sqrt{s} =8\TeV$~\cite{Chatrchyan:2014aqa} and
the ATLAS experiment at $\sqrt{s} =7$, 8, and 13\TeV~\cite{Aad:2012twa,Aad:2016ett,Aaboud:2016yus}.
All measurements are in good agreement with SM predictions.

This paper reports the CMS measurement of the \WZ production
cross section in proton-proton collisions at $\sqrt{s} = 13\TeV$.
The measurement is performed using the leptonic decay modes $\WZ \to \ell\nu\ell'\ell'$, where $\ell, \ell'$ = $\Pe$, $\Pgm$.

\section{The CMS detector}

The CMS detector is described in detail elsewhere~\cite{CMSExperiment}.
The key components for this analysis are summarized here.
The central feature of the CMS apparatus is
a superconducting solenoid of 6\unit{m} internal diameter,
providing a magnetic field of 3.8\unit{T}.
Within the superconducting solenoid volume are a silicon pixel and strip tracker,
a lead tungstate crystal electromagnetic calorimeter (ECAL),
and a brass and scintillator hadron calorimeter,
which provide the pseudorapidity coverage $\abs{ \eta }< 1.479$ in a barrel section
and $1.479 <\abs{ \eta } < 3.0$ in two endcap sections.
Forward calorimeters extend the
coverage to $\abs{ \eta }< 5.0$.
Muons are measured in gas-ionization detectors embedded
in the steel flux-return yoke outside the solenoid.

The first level of the CMS trigger system, composed of custom
hardware processors, is designed to select the most interesting events
in less than 4\mus using information from the calorimeters and muon
detectors.
The high-level-trigger processor farm decreases the event rate from almost $100\unit{kHz}$
to around $1\unit{kHz}$, before data storage.

\section{Data and Monte Carlo samples}

This measurement uses a sample of proton-proton collisions collected
in 2015 at $\sqrt{s} = 13\TeV$.
The integrated luminosity of the sample is 2.3\fbinv.
Several Monte Carlo (MC) event generators are used to simulate the signal and
background processes.

The $\PW\PZ$ signal is generated at NLO in perturbative QCD with
\POWHEG2.0~\cite{Melia:2011tj,Nason:2004rx,Frixione:2007vw,powheg:2010}.
The $\ZZ$ production via $\Pq\Paq$ annihilation is generated at NLO using \POWHEG2.0,
while the $\Pg\Pg \to \ZZ$ process is simulated at leading-order with \MCFM7.0~\cite{Campbell:2011bn}.
The $\cPZ\Pgg$, $\ttbar\text{V}$ ($\ttbar\PW$, $\ttbar\PZ$), $\PQt\PZ$, and triboson events VVV (WWZ, WZZ, ZZZ)  are generated at NLO with \textsc{MadGraph5\_aMC@NLO} ~\cite{mg_amcnlo}.
The $\ZZ$ samples are scaled to the cross section calculated at NNLO for $\Pq\Paq \to \ZZ$ \cite{Cascioli:2014yka} (scaling $k$ factor 1.1)
and at NLO for $\Pg\Pg \to \ZZ$ \cite{Caola:2015psa} (scaling $k$ factor 1.7).
The \PYTHIA8.175~\cite{Sjostrand:2015} program is used for parton showering, hadronization, and
underlying event simulation using the CUETP8M1 tune~\cite{CUETP8M1:2016}.
The NNPDF3.0~\cite{nnpdf} set of parton distribution functions (PDFs)
is used, unless otherwise specified.

For all processes, the detector response is simulated using a detailed
description of the CMS detector, based on the \GEANTfour
package~\cite{GEANT}, and the event reconstruction is performed with
the same algorithms used for data.
The simulated samples include additional interactions per bunch crossing (pileup)
taken from minimum-bias events generated with \PYTHIA.
The simulated events are weighted so that the pileup distribution matches the measured one,
with an average of about 11 pileup interactions per bunch crossing.

\section{Event reconstruction}
\label{sec:eventreconstruction}

Using the information from all CMS subdetectors,
a particle-flow (PF) technique is employed to identify and reconstruct the individual particles
emerging from each collision event
~\cite{CMS-PAS-PFT-09-001,CMS-PAS-PFT-10-001}. The particles  are classified into
mutually exclusive categories: charged hadrons, neutral hadrons, photons, muons, and electrons.

Electrons are reconstructed within the geometrical acceptance $\abs{\eta^{\Pe}} < 2.5$.
The reconstruction combines the information from clusters of energy deposits in the ECAL and the
trajectory in the tracker~\cite{Khachatryan:2015hwa}.
Electron identification relies on the electromagnetic
shower shape and other observables based on tracker and calorimeter information.
The selection criteria depend on the transverse momentum, $\pt$, and $\abs{\eta}$, and on a categorization according to observables
that are sensitive to the amount of bremsstrahlung emitted along the trajectory in the tracker.
Two working points are defined: \textit{tight} and \textit{very tight}.

Muons are reconstructed within $\abs{\eta^{\Pgm}} < 2.4$~\cite{Chatrchyan:2012xi}.
The reconstruction combines the information from both the tracker and the
muon spectrometer.
The muons are selected from among the reconstructed muon track candidates
by applying minimal quality requirements on the track components in the muon system
and by ensuring that muons are associated to small energy deposits in the
calorimeters.
A single \textit{tight} working point is defined.

The electrons and muons are required to originate from the primary vertex,
which is chosen to be the vertex with the highest sum of $\pt^2$ of
its constituent tracks~\cite{Chatrchyan:2014fea}.
For each lepton track
the distance of closest approach to the primary vertex in the transverse
plane, $d_{xy}$, is required to be less than $0.01~(0.07)\unit{cm}$ for electrons in the barrel (endcap) region and
$0.01\unit{cm}$ for muons with $\pt$ less than $20\GeV$ and $0.02\unit{cm}$ for muons with $\pt$ greater than $20\GeV$.
The distance along the beamline, $d_z$, must be less than $0.4\,(0.6)\unit{cm}$ for electrons in the barrel (endcap) and
$0.1\unit{cm}$ for muons. For the \textit{very tight} electron working point, electrons must pass $d_{xy} \leq 0.01\,(0.04)\unit{cm}$
and $d_z \leq 0.05\,(0.4)\unit{cm}$ in the barrel (endcap) region.

Jets are reconstructed using PF objects.
The anti-\kt jet clustering
algorithm~\cite{antikt} with $R=0.4$ is used. The standard method for jet energy
corrections~\cite{cmsJEC}
is applied.
These include corrections to the pileup contribution that keep the jet energy correction and the corresponding uncertainty
almost independent of the number of pileup interactions.
To exclude electrons and muons from the jet sample, the
jets are required to be separated from the identified leptons by
$\Delta R>0.3$. In order to reject
jets coming from pileup collisions (pileup jets), a multivariate-based jet identification
algorithm~\cite{CMS-PAS-JME-13-005} is applied.
This algorithm takes advantage of differences in the shape of energy
deposits in a jet cone between hard-scatter and pileup jets. The jets are required to have $\pt > 20\GeV$ and
$\abs{\eta}<5.0$. To identify the top quark background contribution in its decay to b quarks,
the CSVv2 b tagging algorithm~\cite{Chatrchyan:2012jua} with the \textit{tight} working point
is used~\cite{CMS-PAS-BTV-15-001}. The efficiency for selecting b quark jets is ${\approx} 49 \%$ with
a misidentification probability of ${\approx}4\%$ for c quark jets and ${\approx}0.1\%$ for light quark jets.

The isolation of individual electrons or muons is defined relative to their transverse momentum $\pt^{\ell}$
by summing over the transverse momenta of charged hadrons and neutral particles within a cone with radius
$\Delta R = \sqrt{\smash[b]{(\Delta \eta)^{2} + (\Delta \phi)^{2}}} < 0.4$ around the lepton direction
at the interaction vertex:
\ifthenelse{\boolean{cms@external}}{
\begin{multline}
I^{\ell} = \Bigg( \sum  \pt^\text{charged}\\
+ \text{max}\Big[ 0, \sum \pt^\text{neutral}
+  \sum \pt^{\gamma} - \pt^{\mathrm{PU}}  \Big] \Bigg) \Big/  \pt^{\ell}.
\end{multline}
}{
\begin{equation*}
I^{\ell} = \left( \sum  \pt^\text{charged} + \text{max}\left[ 0, \sum \pt^\text{neutral}
                                 +  \sum \pt^{\gamma} - \pt^{\mathrm{PU}}  \right] \right) /  \pt^{\ell}.
\end{equation*}
}
Here, $\sum  \pt^\text{charged}$ is the scalar sum of the transverse momenta of charged hadrons
originating from the primary vertex.
The $\sum \pt^\text{neutral}$ and $\sum \pt^{\gamma}$ are the
scalar sums of the transverse momenta for neutral hadrons and photons, respectively.
The neutral contribution to the isolation from pileup events,
$\pt^{\mathrm{PU}}$, is estimated differently for electrons and muons.
For electrons,
$\pt^{\mathrm{PU}} \equiv \rho \, A_\text{eff}$,
where the average transverse momentum
flow density $\rho$ is calculated in each event using the ``jet area'' method~\cite{Cacciari:2007fd}, which defines
$\rho$ as the median of the ratio of the jet transverse momentum to the jet area, $\pt^\text{jet}/A_\text{jet}$,
for all pileup jets in the event.
The effective area $ A_\text{eff}$ is the geometric area of the isolation cone times an $\eta$-dependent correction factor
that accounts for the residual dependence of the isolation on the pileup.
For muons, $\pt^{\mathrm{PU}} \equiv 0.5 \sum_i \pt^{\mathrm{PU},i}$,
where $i$ runs over the charged hadrons originating from pileup vertices and
the factor 0.5 corrects for the ratio of charged to neutral particle contributions in the isolation cone.
Electrons are considered isolated
if $I^{\Pe} < 0.08\,(0.07)$ for the barrel (endcap) region, while muons are considered isolated if
$I^{\Pgm} < 0.15$. For the \textit{very tight} electron working point, the electrons must pass
$I^{\Pe} < 0.04\,(0.06)$ for the barrel (endcap) region.

The missing transverse momentum vector \ptvecmiss is defined as
the projection onto the plane perpendicular to the beams of
the negative vector sum of the momenta of all reconstructed PF objects in an event,
corrected for the pileup contribution.
Its magnitude is referred to as \ETmiss.

The overall efficiencies of the reconstruction, identification, and isolation
requirements for the
prompt $\Pe$ or $\Pgm$ are measured in data in several bins of $\pt^{\ell} $ and $ |\eta^{\ell}| $ using a ``tag-and-probe''
technique~\cite{CMS:2011aa} applied to an inclusive sample of \cPZ\ events.
The efficiency for selecting electrons in the ECAL barrel (endcaps) varies from
about 85\% (77\%) at $\pt^{\Pe} \approx 10\GeV$
to about 95\% (89\%) for $\pt^{\Pe} > 20\GeV$.
It is about 85\% in the transition region between the ECAL barrel and endcaps,
$1.44 < \abs{\eta} < 1.57$, averaging over the whole $\pt$ range.
Muons are reconstructed and identified with efficiency above $98\%$ in the
full $\abs{\eta^{\Pgm}} < 2.4$ range. These efficiencies are measured in data
and simulation. The data/MC efficiency ratios are used as scale factors
to correct the simulated event yields.

\section{Event selection}

Collision events are selected by triggers that require the presence of
one or two electrons or muons.
The \pt threshold for the single lepton is 23\,(20)\GeV for the electron (muon) trigger.
For the dilepton triggers, with the same or different flavors, the minimum \pt of the leading and subleading leptons are 17\,(17) and 12\,(8)\GeV
for electrons (muons), respectively.
The trigger efficiency for events within the acceptance of this analysis is greater than 99\%.

A selected event is required to have three lepton candidates $\ell\ell'\ell'$.
The $\ell'\ell'$ pair has two leptons with opposite charge and the same flavor,
as expected for a $\PZ$ boson candidate.
One of the leptons from the $\PZ$ boson candidate is required to have $\pt > 20\GeV$ and the other
$\pt > 10\GeV$.
If more than one combination is possible,
the one with invariant mass closest to the $\cPZ$ boson mass is selected.
The lepton associated with the $\PW$ boson must have $\pt > 20\GeV$. All leptons
must pass the \textit{tight} identification and isolation requirements.
To further reduce the contribution from $\cPZ$+jets in events with an electron
associated with the $\PW$ boson, this electron must pass the requirements of the \textit{very tight} working point.

There must be no other isolated leptons with $\pt > 10\GeV$ in the events.
To reduce contributions from $\ttbar$ events,
the two leptons constituting the $\cPZ$ boson candidate are required to have an invariant mass satisfying
$76 < m_{\ell\ell} < 106\GeV$,
and there must be no jets
with $\pt>20\GeV$ and $\abs{\eta}<2.4$
that pass a b tagging requirement.
The $\WZ$ events are expected to have missing transverse energy consistent with the presence of a neutrino in the final state,
therefore $\ETmiss > 30\GeV$ is required.
The invariant mass of any dilepton
pair must be greater than 4\GeV.
This requirement prevents problems with collinear emission of same-flavor opposite-sign dilepton pairs in theoretical calculations.
The selection is extended to all dilepton pairs at the detector level to reduce backgrounds
from low mass resonances with a negligible effect on signal efficiency.
The trilepton invariant mass, $m_{3\ell}$, is required to be more than 100\GeV
to exclude a region where production of $\PZ$ bosons with final-state radiation
is expected to contribute.

\section{Background estimation}

The background contributions in this analysis are divided into two categories:
background processes with prompt isolated leptons, \eg,\ $\PZ\PZ$, $\cPZ\gamma$, $\ttbar\cPZ$;
and background processes from nonprompt leptons from hadrons decaying to leptons inside jets or
jets  misidentified as isolated leptons, primarily $\cPZ$+jets and $\ttbar$.
The background processes with prompt leptons are estimated from simulation.
The processes with at least one nonprompt lepton are estimated from data.

The major background contributions with nonprompt leptons arise from
the production of $\PZ$ bosons in association with
jets and from $\ttbar$, whereas smaller contributions come from $\PW$ boson production in association with jets and multijet processes.
The nonprompt background contribution is evaluated using the ``\textit{tight}-to-\textit{loose}'' method.
The method
estimates the probability that a \textit{loose} candidate is
misidentified as a \textit{tight} lepton and applies this probability to control regions
with \textit{loose} candidates to estimate the resulting contribution to the signal region.
These \textit{loose} candidates are selected with relaxed lepton identification and isolation requirements.

The misidentification probability is measured from a sample of
dijet events enriched in nonprompt leptons. The sample is selected with one jet
passing the relaxed lepton identification requirements matched to a single lepton trigger,
defined as the probe lepton.
The probe lepton and the second jet must be separated by $\Delta R>1$.
The misidentification ratio for each lepton flavor is defined in bins of lepton $\pt$ and $\eta$
as the ratio of the number of probe leptons that pass the final
isolation and identification requirements to the number of probe leptons that do not pass the
\textit{tight} requirements.
The contamination from $\PW$+jets is suppressed by requiring a transverse mass $m_{\mathrm{T}} < 20\GeV$,
where $m_{\mathrm{T}} = \sqrt{2\ETmiss\pt^\ell(1-\cos(\Delta \phi))}$ and $\Delta\phi$ is the azimuthal
angle between the vectors \ptvecmiss and $\vec{p}_\mathrm{T}^\ell$.
The contamination from $\PZ$ boson events is suppressed by requiring the invariant mass of
each pair of leptons composed of the probe lepton and of any other lepton candidate in the event
to be outside of the window 60--120\GeV.
Contributions from low mass resonances decaying into pairs of leptons are suppressed by requiring the dilepton
mass to be greater than 20\GeV.
The transverse momentum spectrum of the probe lepton in dijet events is different from the spectrum in
$\PZ$ and $\ttbar$ events. We have verified in data that one can make them similar
with a requirement on the
minimum transverse momentum of the second jet
of 20 (35)~\GeV for the dijet events with one probe muon (electron).

A set of control regions with events containing three leptons
is then used to estimate the background from nonprompt leptons.
Zero, one, or two leptons are required to pass the
signal region requirements, while the remaining leptons must
pass the \textit{loose} requirements and fail the signal region requirements.
The misidentification ratio is applied to the \textit{loose}
leptons failing the \textit{tight} identification requirements to estimate the corresponding contribution to the signal region.
The total background is calculated as a sum of contributions from different
regions. This method is validated in nonoverlapping data samples enriched in Drell--Yan and $\ttbar$ contributions.
The Drell--Yan region is defined by inverting the selection requirement in $\ETmiss$ and
the $\ttbar$ region is defined by requiring at least one b-tagged jet and rejecting events with $76 < m_{\ell\ell} < 106\GeV$
while keeping all other requirements for the signal region.
The overall yield predicted with the ``\textit{tight}-to-\textit{loose}'' method agrees with that measured in the control region within 5\%,
with a maximum deviation of 30\% in a single decay channel.
The observed deviations are used as systematic uncertainties in the predicted background yields in the signal region.

\section{Systematic uncertainties}

Systematic uncertainties are less than $1\%$ for the trigger efficiency and 2--4\% for the
lepton identification and isolation requirements, depending on the lepton flavors.
Other systematic uncertainties are related to the use of simulated samples:
1\% for the effects of pileup and  1--2\% for the \ETmiss reconstruction,
which is estimated by varying the energies of the PF objects within their uncertainties.
The uncertainty in the b quark jet content in $\PW\PZ$ events is 2\%
and accounts for differences in b-tagging efficiencies between data and MC as well as differences
in b quark jet content between $\PZ$+jets and $\PW\PZ$+jets events.
The uncertainty in the integrated luminosity of the data sample is 2.7\% \cite{CMS-PAS-LUM-15-001}.
This uncertainty affects both the signal and the simulated portion of the background estimation
and does not affect the background estimation from data;
the total effect of the luminosity uncertainty on the cross section is 3.2\%.

Uncertainties in prompt background sources are estimated from the theoretical uncertainties in the
cross sections. For the $\ZZ$ background the uncertainty is 4\%~\cite{Grazzini:2015hta,Cascioli:2014yka},
and it contributes to the WZ cross section with an uncertainty of 0.4\%.
The uncertainties are
15\% for $\ttbar \text{V}$~\cite{mg_amcnlo,Campbell:2012dh,Garzelli:2012bn} and 6\% for triboson and $\cPZ\Pgg$~\cite{Campbell:2011bn}; their contribution to the uncertainty in the WZ cross section is much
less than 1\%.

The uncertainties in background contributions from both flavors of nonprompt leptons are determined
by combining the uncertainties in the measured values of the misidentification probabilities and the
statistical uncertainties due to the limited number of events in
the control regions. The systematic uncertainty in the misidentification probability is 30\%
for both electrons and muons.
It covers the largest difference observed
between the estimated and measured
numbers of events in data control samples enriched in
$\ttbar$ and Drell--Yan contributions.
The uncertainties are uncorrelated between electrons and muons.
The contribution to the uncertainty in the cross section measurement is 5.4\%\,(3.9\%) from muons (electrons).

Theoretical uncertainties in the $\PW\PZ \to \ell\nu\ell'\ell'$ acceptance are evaluated using \POWHEG
and \MCFM by varying dynamic renormalization and factorization scales independently up and down by a factor of two with respect to the default values
$\mu_R = \mu_F = m_{\PW\cPZ}$ with the condition that $0.5 \leq \mu_R/\mu_F \leq 2$, where $m_{\PW\cPZ}$ is the mass of the WZ system at the generator level.
The uncertainty in the acceptance due to the scale variations
 can be neglected. Phenomenological uncertainties
(PDF${+} \alpha_s$) are estimated
using the CT14~\cite{Dulat:2015mca}, NNPDF3.0, and MMHT2014~\cite{Harland-Lang:2014zoa} PDF sets according to their individual prescriptions.
The largest variation among the sets defines an envelope of about 1\%,
which is taken as the theoretical uncertainty
in the measured cross section.

A summary of each systematic uncertainty and its contribution to the final uncertainty in the
cross section measurement is presented in Table~\ref{tab:systCrossSection}.

\begin{table*}
    \centering
    \topcaption{ The contributions of each systematic uncertainty source to the combined uncertainty in the cross section measurement.
              The integrated luminosity as well as the PDF and scale uncertainties are reported separately in Equations~\ref{eqn:fid} and~\ref{eqn:tot}
              as $\lum$ and $\thy$, respectively,
              while the other uncertainties are combined into a single systematic uncertainty $\syst$. }
    \label{tab:systCrossSection}
    \begin{tabular}{l|c }
\hline
Source of uncertainty                & Uncertainty in the cross section    \\
\hline
Background with nonprompt $\Pgm$ &  5.4\%  \\
Background with nonprompt $\Pe$  &  3.9\%  \\
b tagging                        &  2.1\%  \\
\ETmiss                          &  2.0\%  \\
Electron efficiency              &  1.9\%  \\
Muon efficiency                  &  1.5\%  \\
Pileup                           &  0.8\%  \\
ZZ cross section                 &  0.4\%  \\
$\ttbar \text{V}$ cross section  &  negligible  \\
$\cPZ\gamma$ cross section       &  negligible  \\
VVV cross section                &  negligible  \\
\hline
Integrated luminosity            &  3.2\%  \\
PDF and scales                  &  1.0\%  \\
\hline
     \end{tabular}
\end{table*}

\section{Results}

\begin{figure*}[htbp]
 \begin{center}
\includegraphics[width=0.49\textwidth]{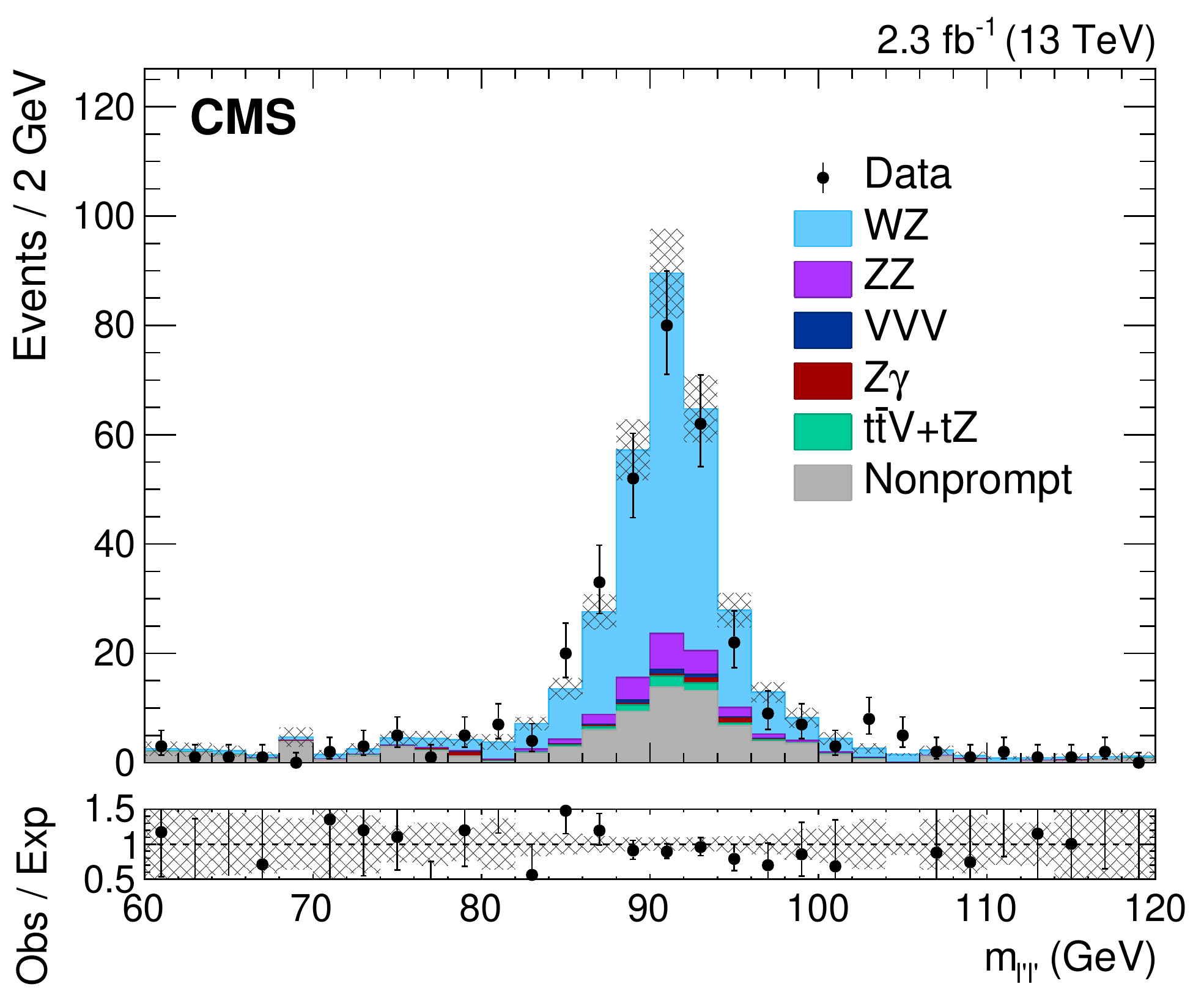}
\includegraphics[width=0.49\textwidth]{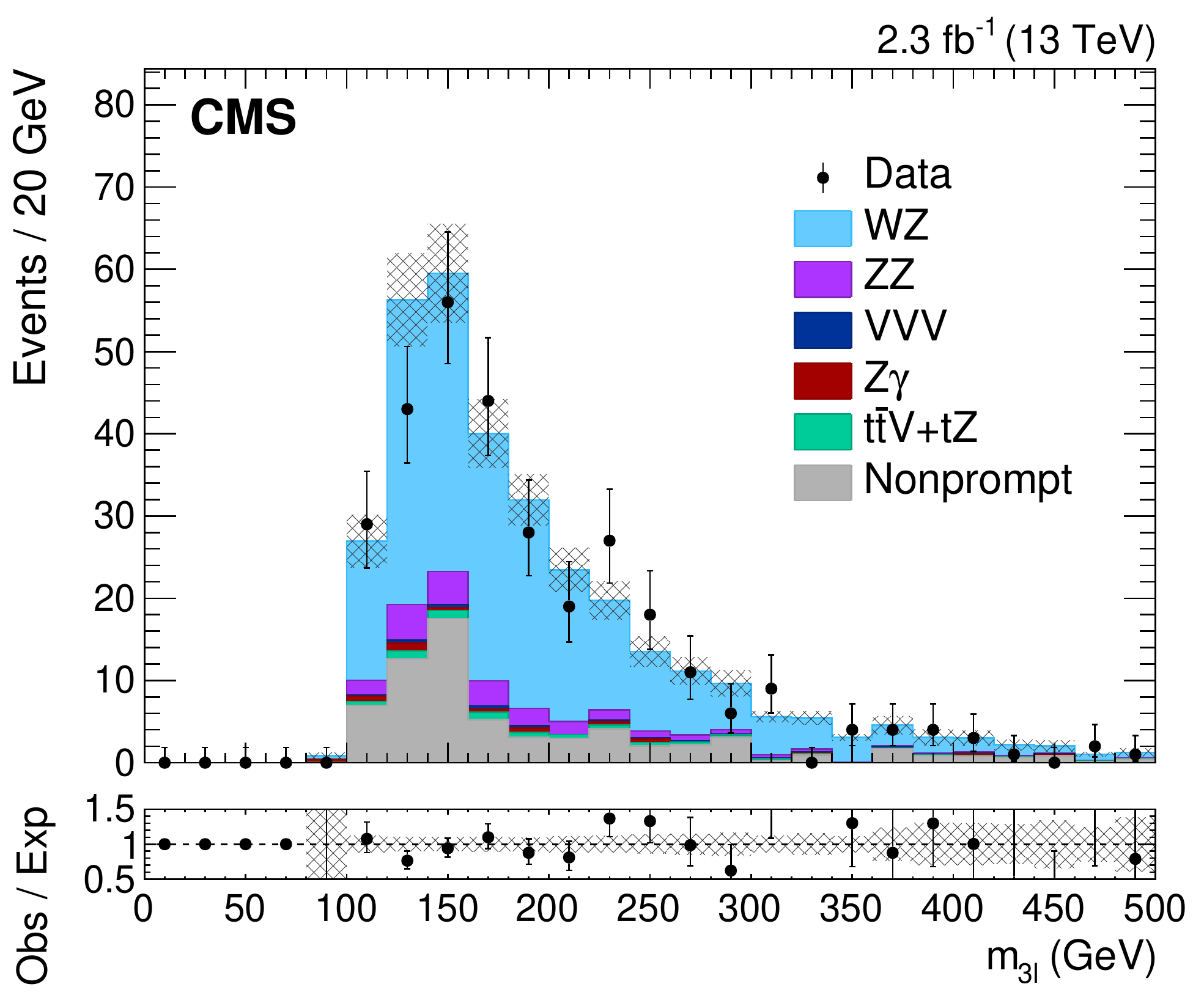}
\includegraphics[width=0.49\textwidth]{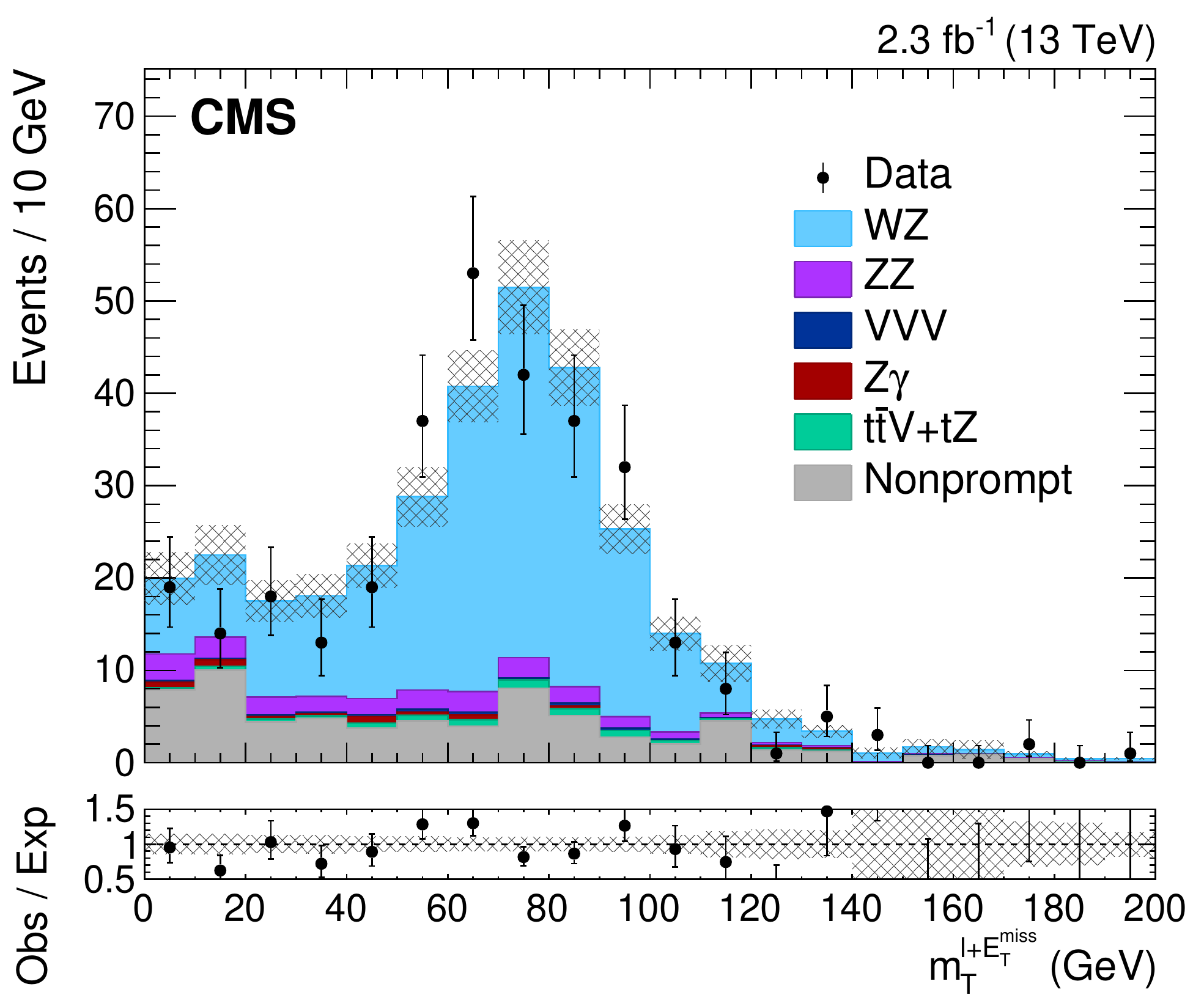}
\includegraphics[width=0.49\textwidth]{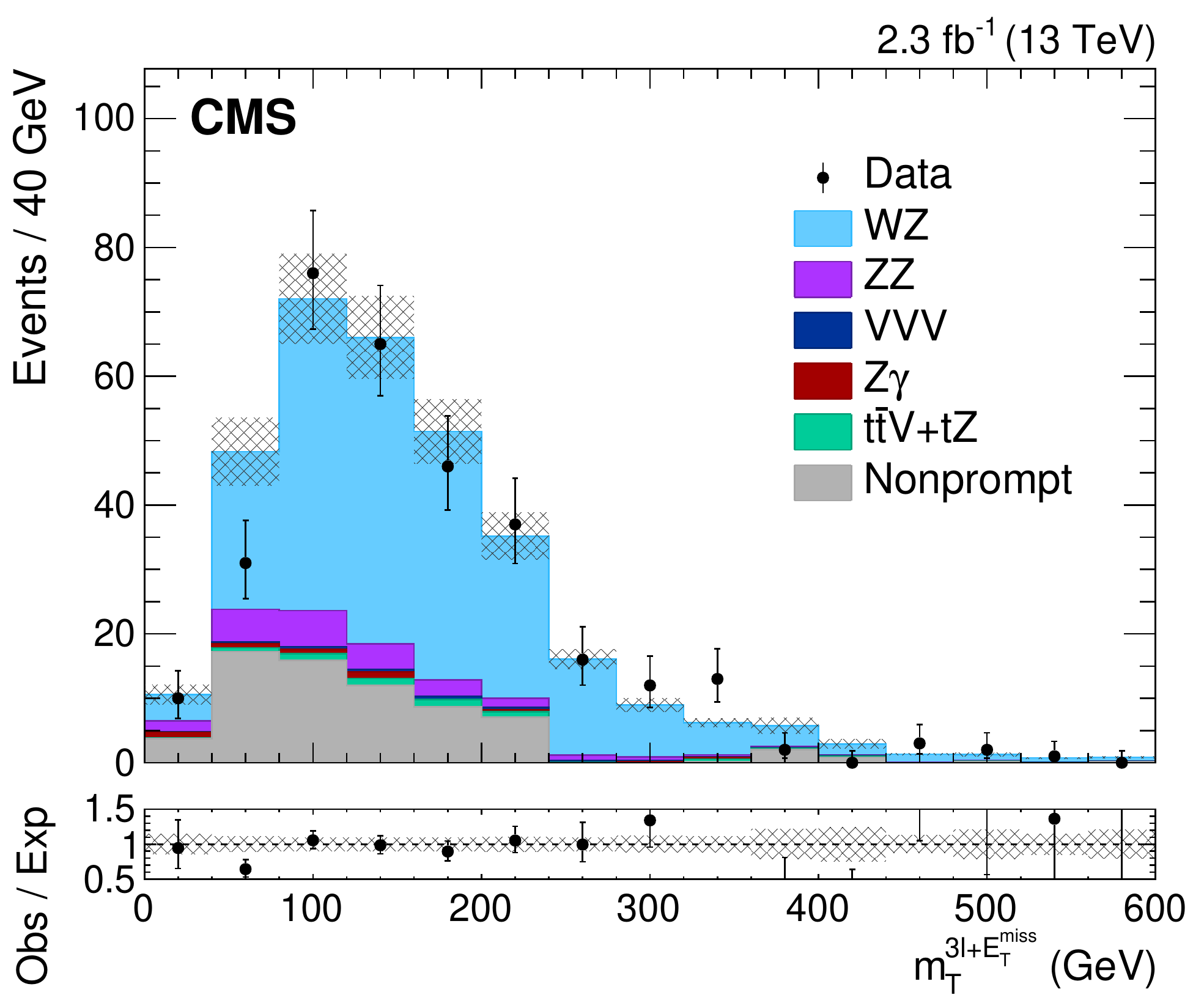}
\caption{
(upper left) Distribution of the reconstructed $\ell'\ell'$ pair mass
 summed for all decay channels with the
$m_{\ell'\ell'}$ selection extended to 60--120\GeV.
(upper right) Distribution of the $\ell\ell'\ell'$ reconstructed mass
 summed for all decay channels with the $m_{3\ell} > 100\GeV$ selection
requirement removed.
(lower left) The transverse mass of the lepton from the $\PW$ boson and the $\ETmiss$ system.
(lower right) The transverse mass of the three leptons and the $\ETmiss$ system.
Solid symbols represent the data with Poisson statistical uncertainties, while histograms represent
the expected WZ signal and  backgrounds.
The shaded band represents the uncertainties in the signal and background estimated yields
and includes systematic, theoretical, and integrated luminosity uncertainties in addition
to the statistical uncertainty.
The background shapes are taken from simulation or
data, as described in the text.
A ratio of the observed (Obs) and expected (Exp) distributions is also included.
\label{fig:MassDistribution}}
 \end{center}
\end{figure*}

\begin{figure*}[htbp]
\centering
\includegraphics[width=0.49\textwidth]{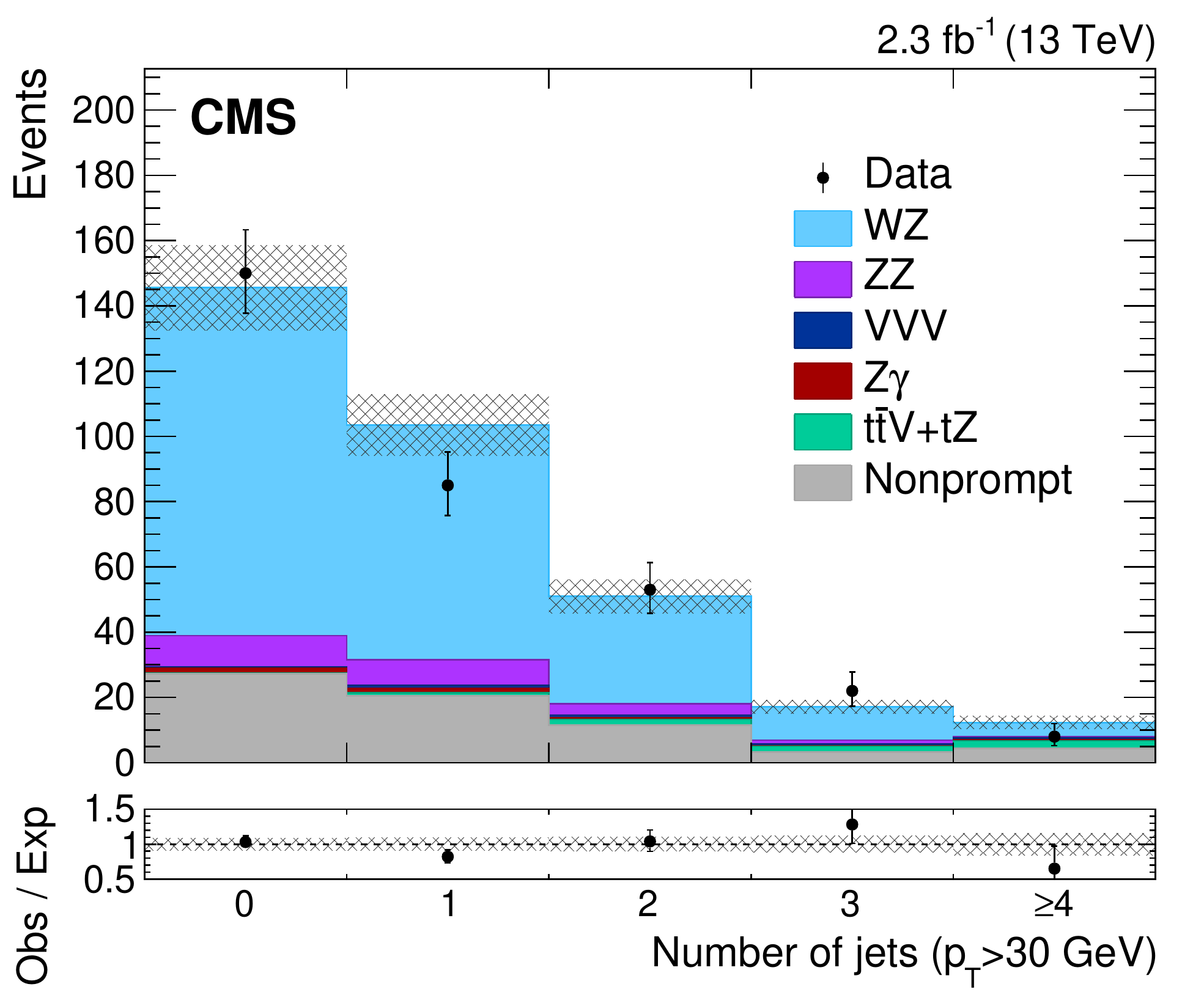}
\includegraphics[width=0.49\textwidth]{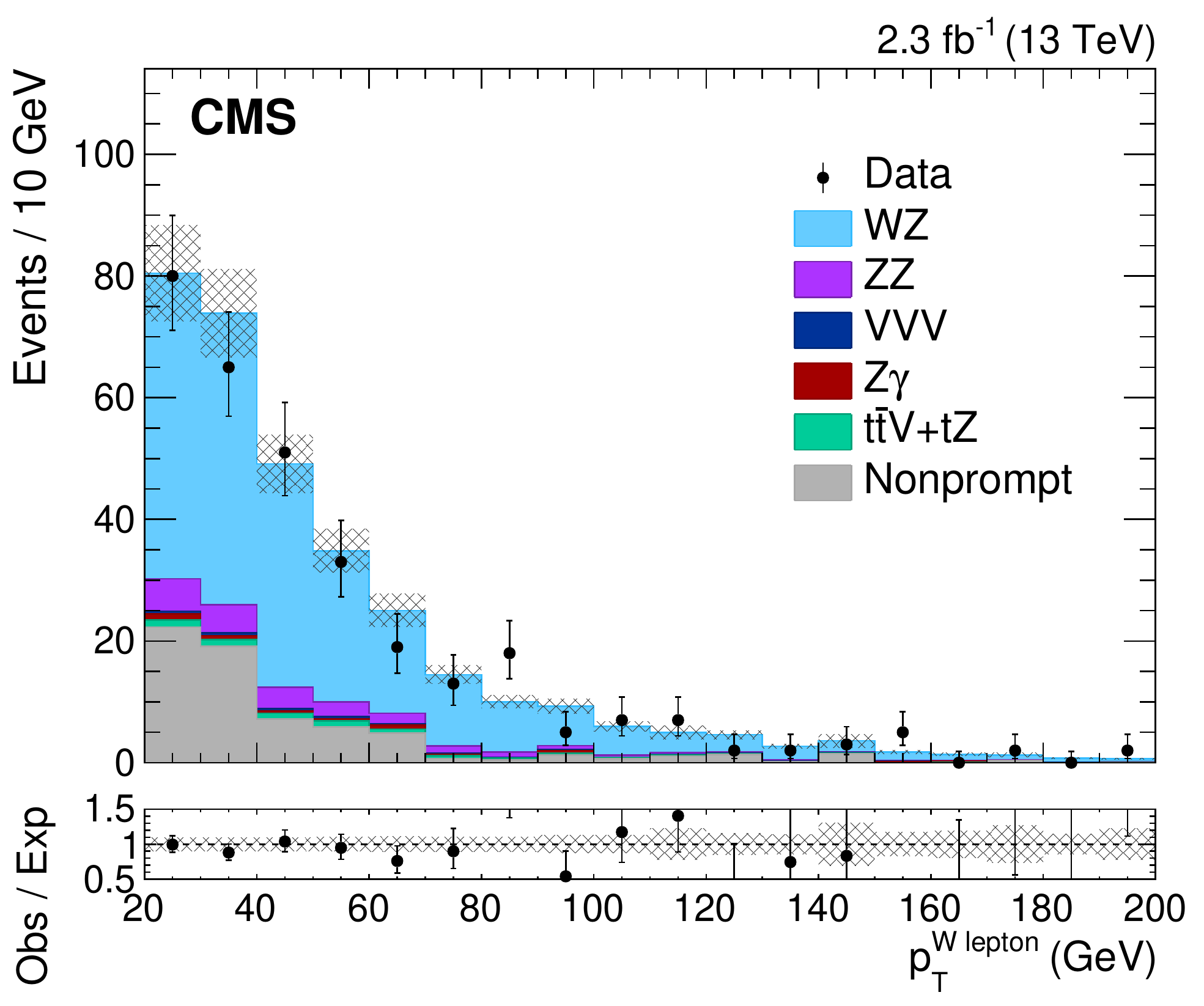}
\includegraphics[width=0.49\textwidth]{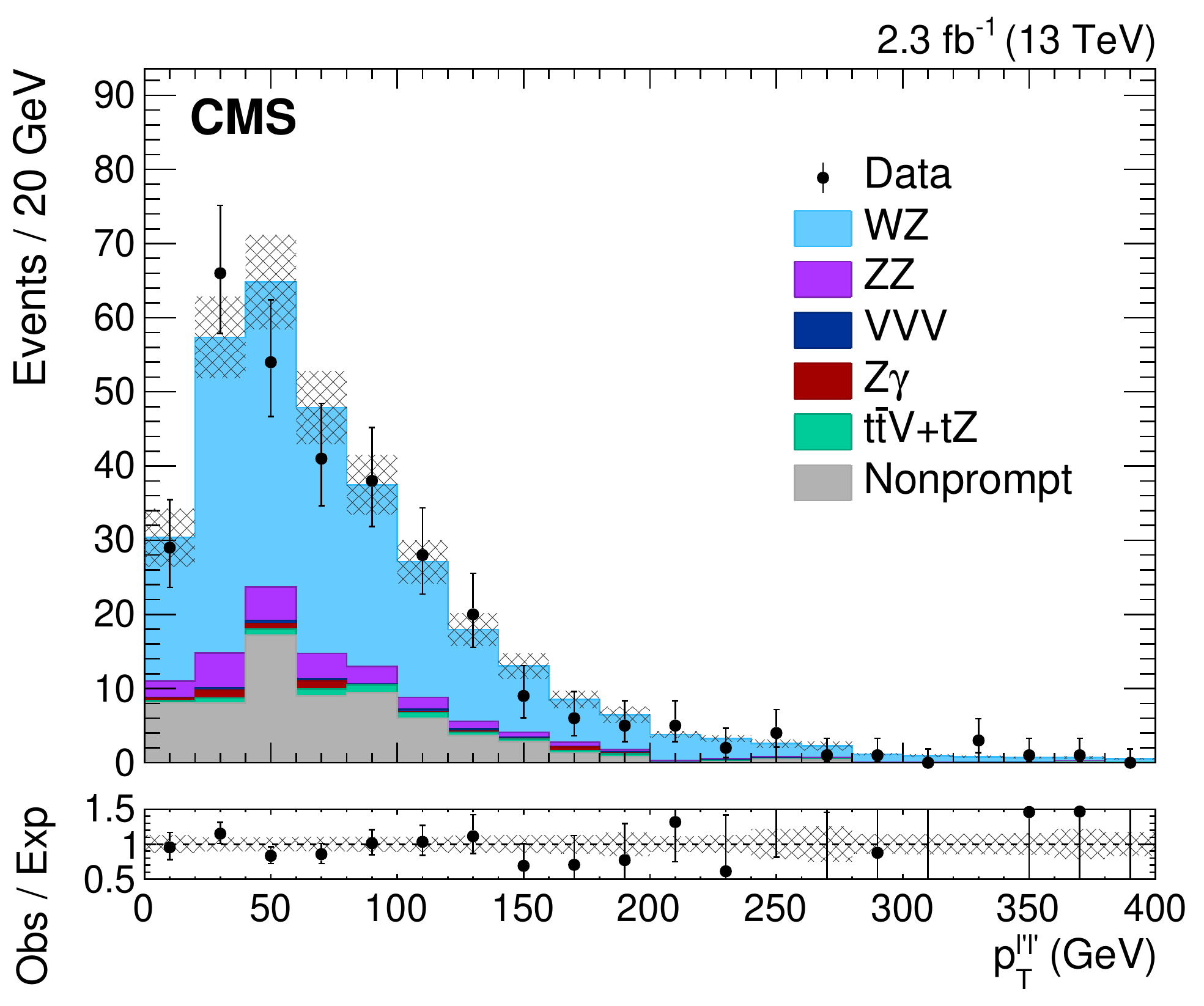}
\includegraphics[width=0.49\textwidth]{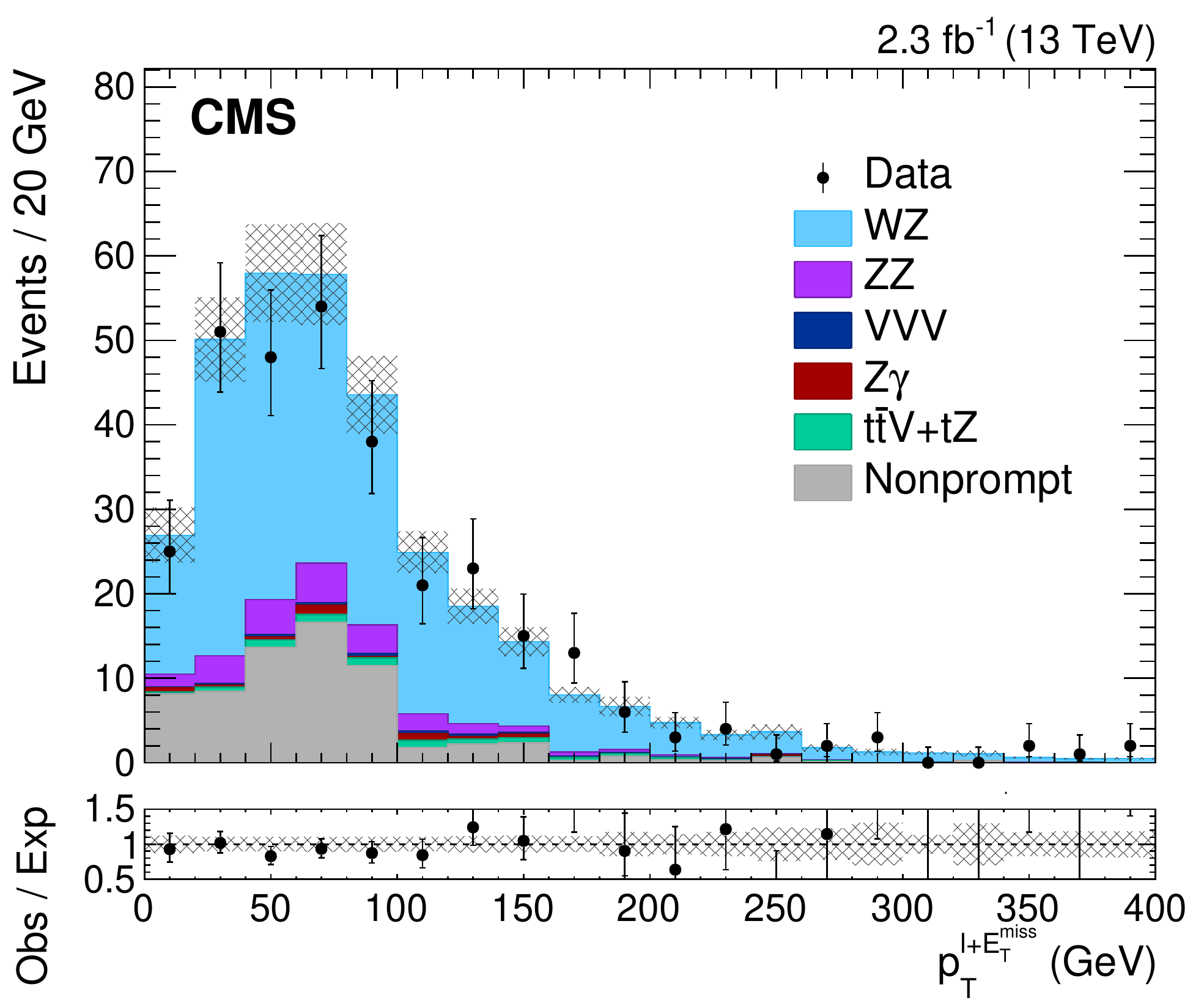}
\caption{
(upper left) Distribution of the number of jets with $\pt > 30\GeV$ in the event.
(upper right) Transverse momentum of the lepton associated with the $\PW$ boson.
(lower left) Transverse momentum of selected $\PZ$ boson candidates.
(lower right) Transverse momentum of selected $\PW$ boson candidates.
Solid symbols represent the data with Poisson statistical uncertainties, while histograms represent
the expected WZ signal and  backgrounds.
The shaded band represents the uncertainties in the signal and background estimated yields
and includes systematic, theoretical, and integrated luminosity uncertainties in addition
to the statistical uncertainty.
The background shapes are taken from simulation or
data, as described in the text.
A ratio of the observed (Obs) and expected (Exp) distributions is also included.
}
   \label{fig:KinematicDistribution}
\end{figure*}

The observed and expected event yields for all decay channels
are summarized in Table~\ref{table:results}.
The invariant mass distributions
for all channels combined
are shown in Fig.~\ref{fig:MassDistribution} and compared to the SM expectations and to the backgrounds
estimated from data.
The two upper plots show distributions for events after all the selection requirements are applied
except the one displayed. The two lower plots show distributions with the full $\PW\cPZ$ selection
requirements.
Kinematic distributions of the selected events are shown in Fig.~\ref{fig:KinematicDistribution}.
Overall, the simulated signal combined with the background contributions are in agreement with the data within uncertainties.

\begin{table*}[htb]
\topcaption{ The expected yields of WZ events and the estimated yields of background events,
consisting of the prompt leptons estimated from simulation
and nonprompt background from data, compared to the number of observed events for each decay channel.
The first uncertainty is statistical and the second is systematic.}
\label{table:results}
\centering
\renewcommand{\arraystretch}{1.5}
\begin{tabular}{l|c|cc|c|c}
\hline
Decay           & Expected                                  &  \multicolumn{2}{c|}{Background}                                                   & Total                                    & Observed     \\[-2mm]
channel         & WZ                                        &  Nonprompt                              & Prompt                                   & expected                                 &              \\
\hline

$\eee         $ & $   35.9 \pm    0.6 ^{ +1.8 }_{ -1.8 }$   & $   10.6 \pm    1.7 ^{ +3.2 }_{ -2.5 }$ & $    6.6 \pm    0.6 ^{ +0.5 }_{ -0.5 }$  & $   53.1 \pm    1.9 ^{ +3.9 }_{ -3.3 }$  & $      49 $  \\
$\eem         $ & $   50.2 \pm    0.8 ^{ +2.4 }_{ -2.4 }$   & $   14.8 \pm    3.6 ^{ +3.9 }_{ -3.0 }$ & $    8.3 \pm    0.5 ^{ +0.6 }_{ -0.6 }$  & $   73.3 \pm    3.7 ^{ +4.8 }_{ -4.1 }$  & $      78 $  \\
$\mme         $ & $   56.0 \pm    0.8 ^{ +2.5 }_{ -2.4 }$   & $   21.5 \pm    3.2 ^{ +5.0 }_{ -3.9 }$ & $    9.3 \pm    0.6 ^{ +0.8 }_{ -0.7 }$  & $   86.8 \pm    3.4 ^{ +5.8 }_{ -4.8 }$  & $      83 $  \\
$\mmm         $ & $   84.0 \pm    1.0 ^{ +3.4 }_{ -3.3 }$   & $   20.0 \pm    4.9 ^{ +6.1 }_{ -4.7 }$ & $   12.4 \pm    0.5 ^{ +0.8 }_{ -0.7 }$  & $  116.3 \pm    5.0 ^{ +7.2 }_{ -6.0 }$  & $     108 $  \\
\hline
{Total}         & $  226   \pm    2   ^{ +10  }_{ -9   }$   & $   67   \pm    7   ^{ +14  }_{ -11  }$ & $   37   \pm    1   ^{ +3   }_{ -2   }$  & $  330   \pm    7   ^{+18   }_{-16   }$  & $     318 $  \\
\hline
\end{tabular}
\end{table*}

The measured yields, corrected for the efficiency of the event selections and the acceptance of the fiducial
phase space, are used to evaluate the WZ production cross section.

The fiducial $\WZ \to \ell\nu\ell'\ell'$ phase space
is defined by the requirement of two leptons from the $\PZ$ boson decay
to have $\pt > 20$ and $10\GeV$, the charged lepton from $\PW$ boson decay to have $\pt > 20\GeV$, all leptons to be within $\abs{\eta} < 2.5$,
$60 < m_{\ell'\ell'} < 120\GeV$, and invariant mass of any same-flavor opposite-sign lepton pair is above 4\GeV. All the leptons are considered before final-state radiation (FSR).
The difference between the cross section calculation
with leptons before FSR and the cross section with ``dressed'' leptons,
which are obtained by summing the lepton momentum and the momenta of
radiated photons within a cone of $\Delta R<0.1$ around the lepton,
is found to be less than $1\%$.

The correction between the fiducial definition and the selection requirements
takes into account the effect
of the \ETmiss requirement, the reduced $m_{\ell'\ell'}$ mass window in the selection with respect to
the fiducial definition, and the requirements of exactly three isolated
leptons and no b-tagged jets in the event.
A small contribution from $\PW\PZ$ events where the $\PW$ or $\PZ$ boson decays
via $\Pgt$ into an electron or muon is considered as signal at the detector level,
but not at the generator level. Thus the correction for $\tau$ lepton decays
is also taken into account in the selection efficiency.

The efficiency of the selection requirements with
respect to the fiducial requirements varies with the channel, from 55\% in $\mmm$ to 25\% in $\eee$.
It includes a 70\% correction for the \ETmiss requirement at the reconstruction level,
a 7\% correction for the contribution from tau decays
and the effects of the lepton identification requirements.
The difference in the $\PZ$ boson mass window definition
at the selection level and in the fiducial definition has a 2\% effect.
The theoretical uncertainties in these corrections are estimated
by checking differences between the various \POWHEG, \MADGRAPH, and \MCFM predictions
and are found to be much less than 1\% so they are neglected in the fiducial cross section measurement.
The major difference between the channels is the tighter identification
and isolation requirements on the electrons.

To include all final states in the cross section calculation,
the number of  expected signal and background events is fitted to
the number of observed events simultaneously in all decay channels.
The likelihood is written as a combination of
individual channel likelihoods for
the signal and background hypotheses.
The statistical
and systematic uncertainties are included as scaling nuisance parameters
and the correlation between different sources of uncertainties across channels is taken into account.

The fiducial $\WZ \to \ell\nu\ell'\ell'$ cross section for
$\pt^{\ell'} > 20, 10\GeV$, $\pt^{\ell} > 20\GeV$, all leptons within $\abs{\eta} < 2.5$,
$60 < m_{\ell'\ell'} < 120\GeV$, and invariant mass of any same-flavor opposite-sign
lepton pair above 4\GeV is
\ifthenelse{\boolean{cms@external}}{
\begin{multline}
  \sigma_{\text{fid}} (\pp \to \PW\PZ \to \ell\nu\ell'\ell') =  258 \pm  21\stat\\{}^{+19}_{-20}\syst \pm  8\lum\unit{fb},
\label{eqn:fid}
\end{multline}
}{
\begin{equation}
  \sigma_{\text{fid}} (\pp \to \PW\PZ \to \ell\nu\ell'\ell') =  258 \pm  21\stat ^{+19}_{-20}\syst \pm  8\lum\unit{fb},
\label{eqn:fid}
\end{equation}
}
corresponding to a total cross section for the range $60 < m_{\ell'\ell'} < 120\GeV$ of
\ifthenelse{\boolean{cms@external}}{
\begin{multline}
  \sigma(\pp \to \PW\PZ) =  39.9 \pm  3.2\stat\\{}^{+2.9}_{-3.1}\syst \pm 0.4\thy \pm  1.3\lum\unit{pb}.
\label{eqn:tot}
\end{multline}
}{
\begin{equation}
  \sigma(\pp \to \PW\PZ) =  39.9 \pm  3.2\stat  ^{+2.9}_{-3.1}\syst \pm 0.4\thy \pm  1.3\lum\unit{pb}.
\label{eqn:tot}
\end{equation}
}

The acceptance of the fiducial
phase space, $(45.0 \pm 0.4)\%$, is calculated with \POWHEG.
The nominal \PZ to dilepton branching fraction $\mathcal{B}(\PZ \to \ell'\ell')$ is $(3.3658 \pm 0.0023)\%$
for each lepton flavor, while for the $\PW$ boson the average branching fraction to each lepton flavor,
$(10.67 \pm 0.16)\%$, is derived from $(10.71 \pm 0.16) \%$ for the electron channel and $(10.63 \pm 0.15)\%$
for the muon channel~\cite{pdg}.

The measured cross sections can be compared to the theoretical values of $274^{+11}_{-8}\,\rm{(scale)} \pm 4\,\rm{(PDF)} \unit{fb}$ for the
fiducial cross section and
$42.3^{+1.4}_{-1.1}\,\text{(scale)} \pm 0.6\,\mathrm{(PDF)}\unit{pb}$ for the total cross section
calculated with \MCFM at NLO with NNPDF3.0 PDFs,
with dynamic renormalization and factorization scales
set to $\mu_R = \mu_F = m_{\PW\cPZ}$.
The uncertainty is obtained by varying the factorization and renormalization scales independently
up and down by a factor of two with the condition that $0.5 \leq \mu_R/\mu_F \leq 2$.
The \MCFM and \POWHEG predicted cross sections agree within
the statistical uncertainties of the generated samples.

The measured total cross section can
also be compared to the theoretical value of $50.0^{+1.1}_{-1.0}\,\text{(scale)}\unit{pb}$,
available at NNLO via \MATRIX~\cite{Grazzini:2016swo} with fixed QCD scales set to
$\mu_R = \mu_F = \frac{1}{2}(m_{\cPZ} + m_{\PW})$ and NNPDF3.0 PDFs.
Uncertainties in this calculation take into account
only renormalization and factorization scale variations.
The variations are done independently with the condition that $0.5 \leq \mu_R/\mu_F \leq 2$.
The values from \MCFM with this scale
choice are $291^{+16}_{-13}\,\text{(scale)} \pm 4\,\mathrm{(PDF)}\unit{fb}$
for the fiducial and $44.9^{+2.2}_{-1.8}\,\text{(scale)} \pm 0.7\,\mathrm{(PDF)}\unit{pb}$ for the total cross sections.

\section{Summary}
The WZ production cross section in proton-proton collisions
at $\sqrt{s} = 13$\TeV
has been measured
with the CMS experiment at the LHC
using a data sample corresponding to an integrated luminosity of 2.3\fbinv.  The measurement is performed in the leptonic decay modes $\WZ\to \ell\nu\ell'\ell'$, where $\ell, \ell'$ = $\Pe$, $\PGm$.
The measured
fiducial $\WZ \to \ell\nu\ell'\ell'$ cross section for two leptons
from the $\PZ$ boson decay with $\pt > 20$ and $10\GeV$,
the charged lepton from the $\PW$ boson decay with $\pt > 20\GeV$,
all leptons within $\abs{\eta} < 2.5$,
and $60 < m_{\ell'\ell'} < 120\GeV$ is
$\sigma_{\text{fid}} (\pp \to \PW\PZ \to \ell\nu\ell'\ell') =  258 \pm  21\stat\,{}^{+19}_{-20}\syst \pm  8\lum\unit{fb}$.
The corresponding total cross section is
$\sigma(\pp \to \PW\PZ) =  39.9 \pm  3.2\stat\,{}^{+2.9}_{-3.1}\syst \pm 0.4\thy \pm  1.3\lum\unit{pb}$
for the dilepton mass range $60 < m_{\ell'\ell'} < 120\GeV$.
For both cross sections, the invariant mass of any same-flavor opposite-sign
lepton pair is required to be above 4\GeV.
This measurement is compared with the theoretical values of
$274^{+11}_{-8}\,\text{(scale)} \pm 4\,\mathrm{(PDF)}\unit{fb}$ for the
fiducial cross section and
$42.3^{+1.4}_{-1.1}\,\text{(scale)} \pm 0.6\,\mathrm{(PDF)}\unit{pb}$ for the total cross section
calculated with \MCFM at NLO with NNPDF3.0 PDFs,
with dynamic renormalization and factorization scales
set to $\mu_R = \mu_F = m_{\PW\cPZ}$, and with the NNLO prediction from \MATRIX.

\begin{acknowledgments}
We thank Massimiliano Grazzini and his collaborators for providing the NNLO
cross section calculations.
We congratulate our colleagues in the CERN accelerator departments for the excellent performance of the LHC and thank the technical and administrative staffs at CERN and at other CMS institutes for their contributions to the success of the CMS effort. In addition, we gratefully acknowledge the computing centres and personnel of the Worldwide LHC Computing Grid for delivering so effectively the computing infrastructure essential to our analyses. Finally, we acknowledge the enduring support for the construction and operation of the LHC and the CMS detector provided by the following funding agencies: BMWFW and FWF (Austria); FNRS and FWO (Belgium); CNPq, CAPES, FAPERJ, and FAPESP (Brazil); MES (Bulgaria); CERN; CAS, MoST, and NSFC (China); COLCIENCIAS (Colombia); MSES and CSF (Croatia); RPF (Cyprus); SENESCYT (Ecuador); MoER, ERC IUT and ERDF (Estonia); Academy of Finland, MEC, and HIP (Finland); CEA and CNRS/IN2P3 (France); BMBF, DFG, and HGF (Germany); GSRT (Greece); OTKA and NIH (Hungary); DAE and DST (India); IPM (Iran); SFI (Ireland); INFN (Italy); MSIP and NRF (Republic of Korea); LAS (Lithuania); MOE and UM (Malaysia); BUAP, CINVESTAV, CONACYT, LNS, SEP, and UASLP-FAI (Mexico); MBIE (New Zealand); PAEC (Pakistan); MSHE and NSC (Poland); FCT (Portugal); JINR (Dubna); MON, RosAtom, RAS and RFBR (Russia); MESTD (Serbia); SEIDI and CPAN (Spain); Swiss Funding Agencies (Switzerland); MST (Taipei); ThEPCenter, IPST, STAR and NSTDA (Thailand); TUBITAK and TAEK (Turkey); NASU and SFFR (Ukraine); STFC (United Kingdom); DOE and NSF (USA).

Individuals have received support from the Marie-Curie programme and the European Research Council and EPLANET (European Union); the Leventis Foundation; the A. P. Sloan Foundation; the Alexander von Humboldt Foundation; the Belgian Federal Science Policy Office; the Fonds pour la Formation \`a la Recherche dans l'Industrie et dans l'Agriculture (FRIA-Belgium); the Agentschap voor Innovatie door Wetenschap en Technologie (IWT-Belgium); the Ministry of Education, Youth and Sports (MEYS) of the Czech Republic; the Council of Science and Industrial Research, India; the HOMING PLUS programme of the Foundation for Polish Science, cofinanced from European Union, Regional Development Fund, the Mobility Plus programme of the Ministry of Science and Higher Education, the National Science Center (Poland), contracts Harmonia 2014/14/M/ST2/00428, Opus 2013/11/B/ST2/04202, 2014/13/B/ST2/02543 and 2014/15/B/ST2/03998, Sonata-bis 2012/07/E/ST2/01406; the Thalis and Aristeia programmes cofinanced by EU-ESF and the Greek NSRF; the National Priorities Research Program by Qatar National Research Fund; the Programa Clar\'in-COFUND del Principado de Asturias; the Rachadapisek Sompot Fund for Postdoctoral Fellowship, Chulalongkorn University and the Chulalongkorn Academic into Its 2nd Century Project Advancement Project (Thailand); and the Welch Foundation, contract C-1845.
\end{acknowledgments}
\bibliography{auto_generated}

\cleardoublepage \appendix\section{The CMS Collaboration \label{app:collab}}\begin{sloppypar}\hyphenpenalty=5000\widowpenalty=500\clubpenalty=5000\textbf{Yerevan Physics Institute,  Yerevan,  Armenia}\\*[0pt]
V.~Khachatryan, A.M.~Sirunyan, A.~Tumasyan
\vskip\cmsinstskip
\textbf{Institut f\"{u}r Hochenergiephysik der OeAW,  Wien,  Austria}\\*[0pt]
W.~Adam, E.~Asilar, T.~Bergauer, J.~Brandstetter, E.~Brondolin, M.~Dragicevic, J.~Er\"{o}, M.~Flechl, M.~Friedl, R.~Fr\"{u}hwirth\cmsAuthorMark{1}, V.M.~Ghete, C.~Hartl, N.~H\"{o}rmann, J.~Hrubec, M.~Jeitler\cmsAuthorMark{1}, A.~K\"{o}nig, I.~Kr\"{a}tschmer, D.~Liko, T.~Matsushita, I.~Mikulec, D.~Rabady, N.~Rad, B.~Rahbaran, H.~Rohringer, J.~Schieck\cmsAuthorMark{1}, J.~Strauss, W.~Treberer-Treberspurg, W.~Waltenberger, C.-E.~Wulz\cmsAuthorMark{1}
% ...
\vskip\cmsinstskip
% ...
1:~~Also at Vienna University of Technology, Vienna, Austria\\
% ...

\end{sloppypar}
\end{document}